\documentclass[12pt,a4paper]{article}

\usepackage{graphicx}
\usepackage{cite}

\newcommand{\figwidth}{13.0cm}
\newcommand{\figheight}{7cm}

\begin{document}

\begin{titlepage}

\begin{flushright}
{\small IFIC/02-48\\FTUV/02-1017} \\[15mm]
\end{flushright}

\begin{center}
{\LARGE\bf Quark-hadron-duality in the charmonium and upsilon system}\\[10mm]

{\normalsize\bf Markus Eidem\"uller} \\[4mm]

{\small\sl  Departament de F\'{\i}sica Te\`orica, IFIC,
           Universitat de Val\`encia -- CSIC,}\\
{\small\sl Apt. Correus 22085, E-46071 Val\`encia, Spain} \\[15mm]
\end{center}

\begin{abstract}
\noindent
In this work we discuss the practical and conceptual issues related to
quark-hadron-duality in heavy-heavy systems. 
Recent measurements in the charmonium region allow a direct test of
quark-hadron-duality. 
We present a formula for non-resonant background production in 
$e^+ e^- \to D{\overline D}$ and extract the resonance parameters of the 
$\psi(3S)-\psi(6S)$. The obtained results are used to investigate 
the upsilon energy range.
\end{abstract}

\vfill


\end{titlepage}

\newpage
\setcounter{page}{1}

\newcommand{\nn}{\nonumber}
\newcommand{\kev}{\mbox{\rm keV}}
\newcommand{\mev}{\mbox{\rm MeV}}
\newcommand{\gev}{\mbox{\rm GeV}}
\newcommand{\eqn}[1]{(\ref{#1})}
\newcommand{\MSb}{{\overline{MS}}}
\newcommand{\ep}{\epsilon}
\newcommand{\IM}{\mbox{\rm Im}}
\newcommand{\Li}{\mbox{\rm Li}_2}
\newcommand\lsim{\mathrel{\rlap{\lower4pt\hbox{\hskip1pt$\sim$}}
    \raise1pt\hbox{$<$}}}
\newcommand\gsim{\mathrel{\rlap{\lower4pt\hbox{\hskip1pt$\sim$}}
    \raise1pt\hbox{$>$}}}

\newcommand{\jhep}[3]{{\it JHEP }{\bf #1} (#2) #3}
\newcommand{\nc}[3]{{\it Nuovo Cim. }{\bf #1} (#2) #3}
\newcommand{\npb}[3]{{\it Nucl. Phys. }{\bf B #1} (#2) #3}
\newcommand{\npps}[3]{{\it Nucl. Phys. }{\bf #1} {\it(Proc. Suppl.)} (#2) #3}
\newcommand{\plb}[3]{{\it Phys. Lett. }{\bf B #1} (#2) #3}
\newcommand{\pr}[3]{{\it Phys. Rev. }{\bf #1} (#2) #3}
\newcommand{\prd}[3]{{\it Phys. Rev. }{\bf D #1} (#2) #3}
\newcommand{\prl}[3]{{\it Phys. Rev. Lett. }{\bf #1} (#2) #3}
\newcommand{\prep}[3]{{\it Phys. Rep. }{\bf #1} (#2) #3}
\newcommand{\rpp}[3]{{\it Rept. Prog. Phys. }{\bf #1} (#2) #3}
\newcommand{\zpc}[3]{{\it Z. Physik }{\bf C #1} (#2) #3}
\newcommand{\epjc}[3]{{\it Eur. Phys. J. }{\bf C #1} (#2) #3}
\newcommand{\sjnp}[3]{{\it Sov. J. Nucl. Phys. }{\bf #1} (#2) #3}
\newcommand{\jetp}[3]{{\it Sov. Phys. JETP }{\bf #1} (#2) #3}
\newcommand{\jetpl}[3]{{\it JETP Lett. }{\bf #1} (#2) #3}
\newcommand{\ijmpa}[3]{{\it Int. J. Mod. Phys. }{\bf A #1} (#2) #3}
\newcommand{\hepph}[1]{{\tt hep-ph/#1}} 
\newcommand{\hepth}[1]{{\tt hep-th/#1}} 
\newcommand{\heplat}[1]{{\tt hep-lat/#1}} 
\newcommand{\hepex}[1]{{\tt hep-ex/#1}}



\section{Introduction}

Quantum Chromodynamics describes the strong interactions of quarks of gluons. 
Nevertheless, these particles are not experimentally detected as the physical
states are formed by hadrons. Few methods link the description in terms of 
QCD parameters
to the properties of the hadronic bound states. Among these are QCD sum rules
\cite{svz:79,rry:85,n:89}, lattice QCD \cite{r:92,mm:94}, chiral perturbation theory
\cite{gl:85,p:95} and the $1/N_C$-expansion \cite{h:74,w:79}.
In this work we focus on the method of QCD sum rules where the notion of
quark-hadron-duality (QHD) plays a dominant role.

The optical theorem provides the 
basis for connecting theoretical and phenomenological quantities.
It relates physical measurable observables like
the cross section for hadron production to theoretical quantities usually
expressed by correlators of two- or three-point functions. In the Euclidean domain 
this correlator can be theoretically
calculated by means of the operator product expansion (OPE) \cite{w:69}.
The leading terms
are given by the perturbative expansion which is supplemented by the condensate
contributions. This can be compared to the corresponding quantity extracted
from experiment and in this way it is possible to extract information about 
the system or the QCD parameters.
One of the limitations of the sum rules already becomes
visible. Approaching from a perturbative side, the correlator does not include
real nonperturbative phenomena. Consequently the analysis must be performed in 
a so-called `sum-rule-window' where the OPE of the
correlator is under control and the system still reacts sensitive
to the hadronic parameters. Furthermore, in practical applications the experimental
spectral density is usually only known for the lowest ground states. To
estimate the missing information on the phenomenological side,
the integral over the experimental spectral density is then assumed to equal
the integral over the theoretical spectral density above a certain threshold
energy $s_0$. This is the assumption of global 
quark-hadron-duality. Though being one of the basic assumptions in QCD sum
rules its range of applicability
has only been scarcely explored. The foundation of QHD was laid in
\cite{pqw:76}. Whereas in semileptonic decays and lepton scattering the
concept of duality is under active investigation, e.g. 
\cite{lu:00,ijmo:01,ci:01,ymmopr:00,papa:01,netal:00},
QHD in the context of QCD sum rules has only recently be reinitiated by Shifman,
see \cite{s:00} and references therein. 

In this paper we will discuss both the practical and conceptual aspects related
to QHD. The main part will focus on the charmonium system where new
measurements from  BES \cite{BES:01} in the region between 3.7 GeV and 4.8 GeV
have improved the experimental situation significantly. Since also the
theoretical spectral density can be calculated this allows a thorough comparison and
a stringent test of QHD. We finally extend these investigations to the upsilon
system. 

The following section is dedicated to an estimate of the uncertainty related
to the use of QHD. After a discussion of the theoretical contributions we give
a description for the threshold parameter $s_0$. The
error on $s_0$ indicates the uncertainty related to the assumption of QHD. In section
3 we investigate the charmonium cross section in more detail. Apart from the
$\psi$-resonances the non-resonant $D$-production has a significant impact on
the cross section. We present a model description for this background and
extract the resonance parameters of the $\psi(3)-\psi(6)$. Section 4
discusses the more conceptual issues since the notion of QHD in
heavy-heavy-systems is far from trivial. The following section concentrates
on the upsilon system. In particular, we give an estimate
for the threshold parameter $s_0$, present a model for the non-resonant
$B$-production and check the validity of the OPE. 
Finally we summarise the results.


\section{Quark-hadron-duality in the charmonium system}

In this work we investigate charm production in $e^+ e^-$-collisions
\begin{equation}
  \label{eq:2.a}
          R_c(s)=
        \frac{\sigma(e^+ e^- \to c\overline{c})}
        {\sigma(e^+ e^- \to \mu^+ \mu^-)}=12\pi\,
        \IM\, \Pi(s+i\ep)\,.
\end{equation}
Via the optical theorem, the experimental cross
section is related to the imaginary part of the correlator $\Pi(s)$ defined by 
\begin{equation}
  \label{eq:2.b}
        \Pi_{\mu\nu}(q^2) = i \int d^4 x \ e^{iqx}\, \langle
        T\{j_\mu(x) j_\nu^\dagger(0)\}\rangle
          = (q_\mu q_\nu-g_{\mu\nu}q^2)\,\Pi(q^2)\,.
\end{equation}
The charm vector current is given by $j_\mu(x)=e_c(\overline{c}\gamma_\mu c)(x)$
where $e_c$ represents the electric charge of the charm quark.

In principle one can calculate $\Pi(q^2)$ perturbatively, take the imaginary
part and compare it to the measured cross section. However, a perturbative
calculation of $\Pi(q^2)$ is valid only in the Euclidean domain. An analytic
continuation to the Minkowski region neglects terms which are small in
Euclidean but can become important in Minkowski. Thus, with the assumption of
global QHD, only smeared quantities can be compared. A further complication
arises in the theoretical calculation of $\Pi(q^2)$. In the deep Euclidean
domain the perturbative expansion works well. However, usually one is
interested in a region closer to threshold. Here the Coulomb-like behaviour of
the charmonium system shows up and the theoretical expansion converges
badly. These large terms can be resummed with the help of the theory of 
non-relativistic QCD (NRQCD) \cite{cl:86,bbl:95}. 
Since we will employ the
theoretical prediction for $\Pi(q^2)$ and the theoretical spectral density in
this and the following chapters, first we briefly discuss these contributions
and then return to a discussion of QHD. 

\begin{figure}
\begin{center}
\vspace{5mm}
\includegraphics[width=\figwidth, height=\figheight]{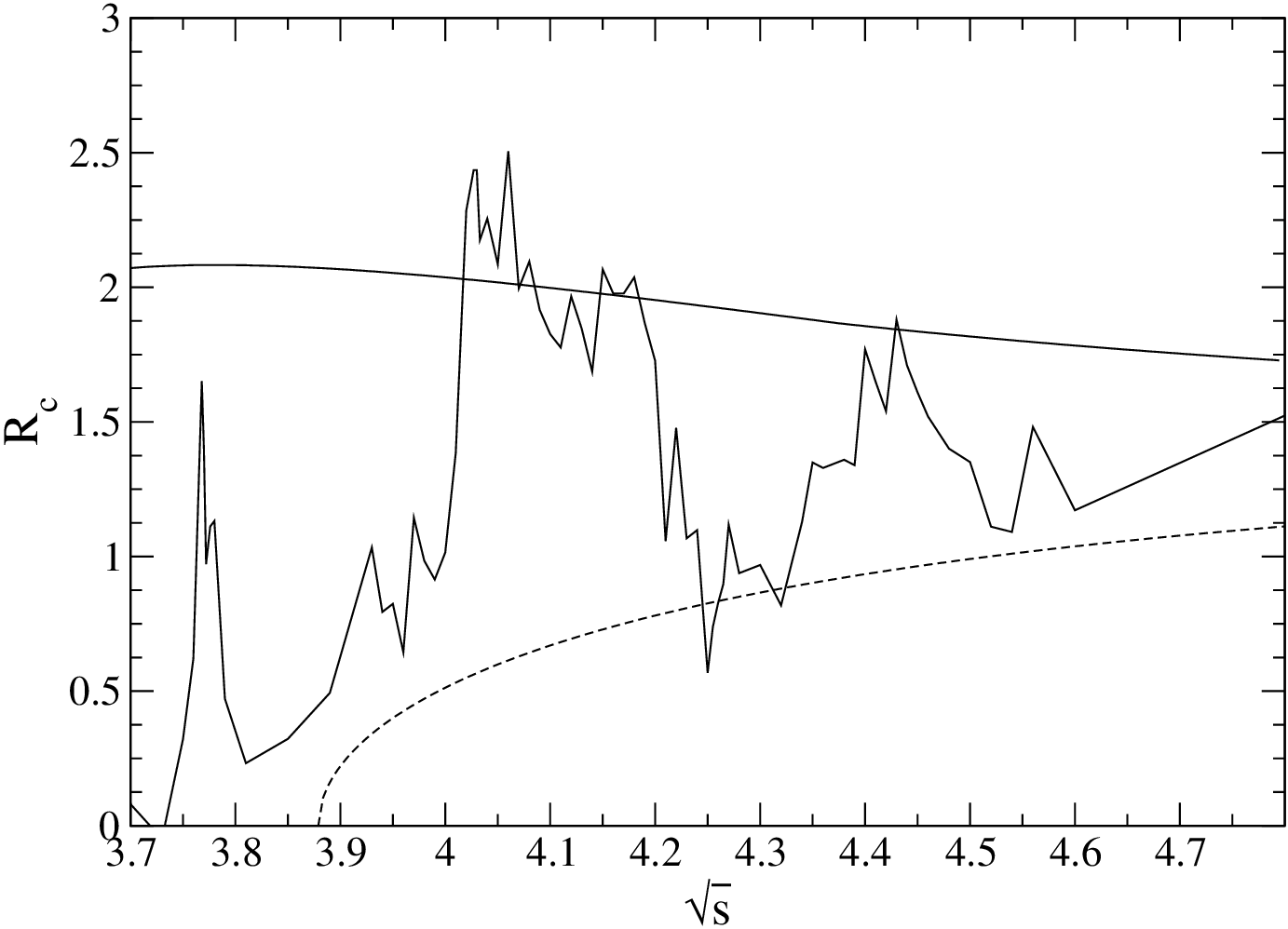}
\caption{\label{fig:2.a}
Cross section from BES with the perturbative spectral density (solid)
and the background contribution (dashed).}
\end{center}
\end{figure}

The theory of NRQCD provides a consistent framework
to treat the problem of heavy quark-antiquark production close to
threshold.
The contributions can be described by a nonrelativistic Schr\"odinger
equation and systematically calculated in time-independent perturbation
theory.
The correlator is expressed in terms of a Green's function
$G(k)=G(0,0,k)$ \cite{ps:91,ht:98,pp:99}:
\begin{equation}
  \label{eq:2.c}
  \Pi(s)=\frac{N_c}{2M_c^2}\left(C_h(\alpha_s)G(k)+\frac{4k^2}{3M_c^2}G_C(k)\right)\,,
\end{equation}
where $N_c$ is the number of colours, $k=\sqrt{M_c^2-s/4}$ and
$M_c$ represents the pole mass. The constant $C_h(\alpha_s)$ is a perturbative coefficient
needed for the matching between the full and the nonrelativistic
theory. The contributions from NRQCD are summarised in the potential.
The Green's function obeys the corresponding Schr\"odinger equation
\begin{equation}
  \label{eq:2.d}
  \Bigg( -\frac{\Delta_x}{M_c}+V_C(x)+\Delta V(x)
  +\frac{k^2}{M_c}\Bigg)
  G({\bf x},{\bf y},k)=\delta^{(3)}({\bf x}-{\bf y}) \,.
\end{equation}
Here $V_C(x)=-C_F\alpha_s/|{\bf x}|$ represents the Coulomb potential
and $\Delta V(x)$ contains the NLO and NNLO corrections.
Details on the solution of this equation can be found in \cite{pp:99}.
The Green's function contains pole contributions below threshold
and a continuum above threshold. 
In order to construct the theoretical continuum spectral density for the
full energy range it is not sufficient to use the spectral density
from NRQCD which is only valid for low velocities. In addition, one
must include the results from perturbation theory which gives  $R(s)$
at large velocities. At intermediate velocities one can perform a 
matching between both regimes. In \cite{e:02} this procedure is 
described in detail. The resulting theoretical spectral density is shown
as a solid line in fig. \ref{fig:2.a} and we denote the spectral density 
by  $R^{theo}(s)$.

This spectral density is supposed to give a good approximation to the 
experimental spectral density at high energies. Decreasing the energy,
one approaches the resonance region. Here $R^{theo}(s)$ will fail
to reproduce the resonances. But naively assuming QHD, a smearing 
of $R^{exp}(s)$ over
a `sufficiently large' energy range should give a good approximation
to the same smearing with $R^{theo}(s)$.
In fact, the notion of QHD is more subtle and the naive expectation
is not correct. The optical theorem relates
the two representations by a dispersion relation which includes all values
of the energy. So only the complete phenomenological result can be compared
to the full theoretical result which also includes the poles of the 
Green's function. In section 4 we 
discuss the concept of QHD in more detail and present quantitative
estimates for the individual contributions.
In fig. \ref{fig:2.a} we have plotted $R^{theo}(s)$
and the measured cross section in the energy range between
$3.7\ \gev<\sqrt{s}<4.8\ \gev$.
At these energies $R^{theo}(s)$ clearly lies above the data points. 
On the other hand, the contribution from the theoretical poles turns out to be
smaller than the contribution from the lowest $\psi$-states.

To test the accuracy of QHD from a comparison of the theoretical and 
phenomenological spectral densities we choose the moments 
\begin{equation}
  \label{eq:2.e}
	W_n(\Delta,s_0,s_1)=\int_{s_0}^{s_1} ds\ 
	\frac{R(s)}{(s+\Delta)^{n+1}}\,.
\end{equation}
The weight function $w(s)=1/(s+\Delta)^{n+1}$ 
corresponds to the one usually used in
the moment sum rules to extract the quark masses. Another popular
weight function is $w_u(s)=e^{-s/u}$ which is used in the 
Borel sum rules. Since the analysis and the results are very similar 
in both cases we will not perform an independent analysis for the 
Borel sum rules in this work. 
$\Delta$ and $n$ are free parameters which can be used to move the sum rules
to a region convenient for the analysis. Large $\Delta$ will improve the
perturbative expansion while for small or even negative $\Delta$ the analysis
will react very sensitive to the bound states. Small $n$ will result in a
relative flat weight function and high $n$ put the emphasis on the low
energy region.
As will be discussed in chapter 4, $s_1$ should be taken to infinity to
establish a precise relation between the phenomenological and theoretical part
where they are connected by an Euclidean quantity. However, in practice the
cross section is only measured up to a certain energy so above this energy one
has to rely on the theoretical prediction for $R(s)$. When comparing both parts
the integral above this energy is then equal.

The experimental cross section can be interpreted
as non-resonant background production of $D^{(\star)}$-mesons
and resonances of Breit-Wigner form. For a comparison to $R^{theo}$ 
it is interesting to separate these two contributions. In the next
section we will give a model description of the background production
which has been plotted as a dashed line in fig. \ref{fig:2.a} 
and discuss the charmonium cross section in more detail.
In this section we focus our attention on the most important issue
in actual sum rule calculations and the basic question of QHD: 
how well can the experimental cross section be approximated
by  the theoretical spectral density and how large is the error on 
$W_n(\Delta,s_0,s_1)$? 
The BES data have been given directly in terms of $R^{\rm BES}(s)$ and we will
use these values as our reference for the experimental cross section.
From the measured spectral density the light quark contribution must
be subtracted. At these energies the light quarks can safely assumed to
be massless and the high energy approximation \cite{c:97} provides a good
description.
Apart from BES, the resonance properties of the $\psi$-states have been
extracted in \cite{betal:78,setal:80} and of the $\Upsilon$-states 
in \cite{b:85,l:85}. 

In order to test a typical problem in the sum rules
let us now assume that the only information we had from the experimental side 
were the resonance properties of the first $p$ bound states.
With the assumption of QHD, the contribution
from the higher states is then given by the integration of the
theoretical spectral density $R^{theo}(s)$ above a threshold $s_0$.
The above question can then be formulated in a different way:
what value of  $s_0$ must be taken? 

Since the QCD sum rules are used to extract the heavy quark masses, the choice of
$s_0$ influences the central value of the masses. Furthermore the uncertainty in
$s_0$ translates directly to the error of the masses and it is therefore
important to have a reliable estimate of this uncertainty. There is no rigorous
justification for a particular choice of $s_0$. As a heuristic rule it is
usually assumed that $s_0$ should be given by about 250 MeV above the 
highest included resonance. In
this section we want to check if - and to what extend - this rule is valid. It will
turn out that particular care must given to a possible background
contribution. In chapter 5 we will then apply the results obtained in this
section to estimate $s_0$ in the upsilon system where no experimental
information is available to fix this parameter.

To determine $s_0$ in the charmonium system we now compare the quantities
\begin{eqnarray}
  \label{eq:2.f}
	\lefteqn{\frac{9\pi}{\alpha^2}\sum_{k=1}^2 
	\frac{\Gamma_{ee,k}\,E_k}{(E_k^2+\Delta)^{n+1}}
	+\int_{3.7^2}^{4.8^2} ds\ \frac{R^{\rm{BES}}(s)}{(s+\Delta)^{n+1}}}\nn\\
	&&=\frac{9\pi}{\alpha^2}\sum_{k=1}^p 
	\frac{\Gamma_{ee,k}\,E_k}{(E_k^2+\Delta)^{n+1}}
	+\int_{s_0^{(p)}}^{4.8^2} ds\ \frac{R^{theo}(s)}{(s+\Delta)^{n+1}}\,,
\end{eqnarray}
where $\alpha$ is the electromagnetic fine structure constant,
$\Gamma_{ee,k}$ is the partial decay width into
$e^+e^-$ and $E_k$ is the mass of the $k^{th}$ resonance. We have used the
narrow-width approximation for the resonances. The sum over the resonances in
the first line extends only over the first 2 resonances since the other known
resonances $\psi(3)-\psi(6)$ are included in $R^{\rm{BES}}(s)$.
The upper limit of the integration is taken to infinity, $s_1\to \infty$.
Since the BES data have only been measured up to 4.8 GeV we assume QHD above
this energy so the integral from 4.8 GeV to infinity is identical on
both sides and drops out. 
The first line represents the `exact' result from the data. The second line
is a typical phenomenological approximation using the assumptions of QHD in
which $s_0$ is left as a free parameter. 
As mentioned above, it is usually assumed that $s_0$ should be given
by the mass of the highest resonance $E_p$ plus about 250 MeV.
In table \ref{tab:2.a} we have listed $s_0$ obtained from eq. \eqn{eq:2.f}
for different values of $n$ and $\Delta$.
\begin{table}
\begin{center}
\begin{tabular}{|c|c||c|c|c|c|c|c|}\hline
\multicolumn{8}{|c|}{$\Delta=0$} \\ \hline
\multicolumn{2}{|c||}{$p$} & 1 & 2 & 3 & 4 & 5 & 6 \\
\multicolumn{2}{|c||}{$E_p/[{\rm GeV}]$} & 3.097 & 3.686 & 3.770 
& 4.040 & 4.159 & 4.415 \\ \hline
& $n=0$ & 3.78 & 4.07 & 4.10 & 4.21 & 4.31 & 4.38  \\
$\sqrt{s_0}/[\gev]$ & $n=3$ & 3.65 & 3.99 & 4.03 & 4.14 & 4.25 & 4.30  \\
& $n=7$ & 3.59 & 3.92 & 3.99 & 4.08 & 4.17 & 4.20  \\ \hline \hline
\multicolumn{8}{|c|}{$\Delta=(4\ \gev)^2$} \\ \hline
\multicolumn{2}{|c||}{$p$} & 1 & 2 & 3 & 4 & 5 & 6 \\ \hline
& $n=0$ & 3.82 & 4.09 & 4.12 & 4.22 & 4.32 & 4.39  \\
$\sqrt{s_0}/[\gev]$ & $n=3$ & 3.72 & 4.03 & 4.07 & 4.18 & 4.29 & 4.35  \\
& $n=7$ & 3.65 & 3.98 & 4.03 & 4.13 & 4.24 & 4.30  \\ \hline
\end{tabular}
\caption{\label{tab:2.a}  $s_0$ for different number of
poles $p$ and different values of $n$ and $\Delta$ from eq. \eqn{eq:2.f}.}
\end{center}
\end{table}

Let us first look on the behaviour of  $s_0$ on $n$ and $\Delta$.
We see that  $s_0$ depends on the choice of these
parameters. Since $R^{theo}$ lies above the experimental
cross section, larger values of $n$ will lower  $s_0$. The difference
between the largest ($n=0$) and smallest ($n=7$) value varies between
$\Delta \sqrt{s_0}=190\ \mev$ for $p=1$ and $\Delta \sqrt{s_0}=110\ \mev$ for 
$p=3$ at $\Delta=0$. At  $\Delta=(4\ \gev)^2$ the analysis is in
a more perturbative region. Therefore one expects less impact of $n$ on
$s_0$, but still the difference remains sizeable:
from $\Delta \sqrt{s_0}=170\ \mev$ for $p=1$ to $\Delta \sqrt{s_0}=80\ \mev$ for 
$p=5$. This is a remarkable result for applications of the sum rules.
In analyses where the threshold  $s_0$ has an important impact
on the quantity one would like to extract, the change of  $s_0$ with
$n$ and $\Delta$ might influence the final result.

Furthermore we note that the rule `highest resonance plus  250 MeV' is strongly
violated. Taking only the lowest pole, $p=1$, this is no surprise.
Since the first two poles are very dominant on the experimental side,
one cannot hope to give a good description of these poles by
the perturbative spectral density without taking into account the
pole contributions from the Green's function. Also the reason for
the violation of the rule for the higher states is clear: 
Using only the resonance parameters $E_k$ and $\Gamma_{ee,k}$
in the second line of eq. \eqn{eq:2.f} we have neglected the 
non-resonant $D^{(*)}$-production. There are two ways
to estimate this background contribution. The first one is to give
a model description for the background as has been depicted in fig. \ref{fig:2.a}.
In a second approach one could assume that QHD above the third pole
already represents a reasonable description. In this case the phenomenological
part is given by `3 poles plus $R^{theo}$ from a threshold of 250 MeV above the
$3^{rd}$ pole'. Without background this description also applies to $p$ poles
instead of three. Subtracting these two descriptions should therefore give an
estimate of the background contribution.

The drawback of both methods is clear: in the first one the result
is model dependent where in the second one it was assumed
that QHD could already be used for states with $p>3$.
To include the background contribution one could either add this 
background explicitly to the second part of eq. \eqn{eq:2.f}
and use $\sqrt{s_0}=E_p +  250\ \mev$ or stick to eq. \eqn{eq:2.f}
and lower $s_0$ by the appropriate value. Since we want to compare
the results with table \ref{tab:2.a}, we use the second method.
$s_0$ is then determined from the equation 
$\int_{s_0}^{s_1}ds\,R^{theo}(s)/(s+\Delta)^{n+1}=\mbox{`background'}$ where 
$\sqrt{s_1}=E_p+250\ \mev$ and the estimate of the background ranges up to
$s_1$. The results for $s_0$ are shown in table \ref{tab:2.b}.
In the first row the background has been estimated by  the model
description from fig. \ref{fig:2.a} and the second row in
parentheses shows the result assuming QHD already for $p>3$.
\begin{table}
\begin{center}
\begin{tabular}{|c|c||c|c|c|c|c|c|}\hline
\multicolumn{8}{|c|}{$\Delta=0$} \\ \hline
\multicolumn{2}{|c||}{$p$} & 1 & 2 & 3 & 4 & 5 & 6 \\ \hline
& $n=0$ & 3.35 & 3.93 & 3.99 & 4.16 & 4.22 & 4.32  \\
& & (3.35) & (3.94) & (4.02) & (4.12) & (4.22) & (4.29)  \\
$\sqrt{s_0}/[\gev]$ & $n=3$ & 3.35 & 3.93 & 3.99 & 4.14 & 4.18 & 4.25  \\
& & (3.35) & (3.94) & (4.02) & (4.12) & (4.23) & (4.28)  \\
& $n=7$ & 3.35 & 3.93 & 3.99 & 4.11 & 4.14 & 4.17  \\
& & (3.35) & (3.94) & (4.02) & (4.13) & (4.24) & (4.28)  \\ \hline \hline
\multicolumn{8}{|c|}{$\Delta=(4\ \gev)^2$} \\ \hline
\multicolumn{2}{|c||}{$p$} & 1 & 2 & 3 & 4 & 5 & 6 \\ \hline
& $n=0$ & 3.35 & 3.93  & 3.99 & 4.16 & 4.22 & 4.33   \\
&  & (3.35) & (3.94)  & (4.02) & (4.12) & (4.22) & (4.29)   \\
$\sqrt{s_0}/[\gev]$ & $n=3$ & 3.35 & 3.93 & 3.99 & 4.15 & 4.21 & 4.29  \\
& & (3.35) & (3.94) & (4.02) & (4.12) & (4.22) & (4.28)  \\
& $n=7$ & 3.35 & 3.93 & 3.99 & 4.14 & 4.18 & 4.24  \\
& & (3.35) & (3.94) & (4.02) & (4.12) & (4.23) & (4.28) \\ \hline
\end{tabular}
\caption{\label{tab:2.b}  $s_0$ for different number of
poles $p$ and different values of $n$ and $\Delta$ obtained from
the description `$\sqrt{s_0}=E_p$ plus 250 MeV minus background'. The numbers
in the first row are with a background from fig. \ref{fig:2.a}
and the numbers in parentheses with the second estimate of the background
as described in the text.}
\end{center}
\end{table}

Let us now compare the `exact' result for $s_0$ from table \ref{tab:2.a} which
is needed for a correct description of the experimental moments
to the QHD-based estimate of $s_0$ in table  \ref{tab:2.b}.
For $p=1$ QHD cannot reproduce the correct value. 
This is expected since the theoretical
spectral density will not give a good description for the second
resonance. 
At $p=2,3$ the background has almost no
influence on $s_0$ so it essentially lies 250 MeV above the resonance.
For a relatively steep weight function, $n=7$, the results are similar
to table \ref{tab:2.a},
but for $n=0$ they differ up to 160 MeV. The estimates for $s_0$
agree well for $p=4$ and for $p=5,6$ they are a bit lower. In general,
we see that the assumption of QHD with the corresponding choice for
$s_0$ gives a reasonable description of the experimental moments, at least  
for the higher poles.
It is interesting to note that in table \ref{tab:2.b} the change of 
$s_0$ with $n$ and $\Delta$
is relatively small. With the background model it is smaller than in table
\ref{fig:2.a} especially for a small number of poles $p$. In the second approach  
with QHD assumption for $p>3$ the value of $s_0$ remains almost constant. 
So this variation can easily be underestimated.

To summarise, we have estimated the uncertainty connected with the use of QHD.
In the charmonium system QHD
represents a reasonable good approximation if at least the first two poles
are added explicitly to the phenomenological side. The moments can then
be determined with the description 
`poles with resonance parameters plus background plus theoretical spectral
density above $s_0$'. The threshold $s_0$ should be given by the energy of the highest 
pole $E_p$ plus 250-300 MeV. Estimating the uncertainty on $s_0$ one should
take into account that the variation with $n$ and $\Delta$ can easily
amount to 100 MeV. In addition one should allow a variation of 100 MeV
from its value `$E_p$ plus 250-300 MeV'. So we conclude that
a reasonable error estimate for $s_0$ is given by 
$\Delta \sqrt{s_0}= 200\ \mev$ around its central value.


\section{Charmonium cross section}

In this section we investigate the charmonium cross section in more detail.
The first two resonances are dominating the cross section clearly, 
the $J/\psi$ at 3.097 GeV with a partial decay width of
$\Gamma_{ee}=5.26$ keV and the $\psi'$ at 3.686 GeV with $\Gamma_{ee}=2.19$ keV.
Since both resonances lie below open $D$-production
their total widths are small, $\Gamma_{tot}=87$ keV and 
$\Gamma_{tot}=300$ keV \cite{pdg:02} (264 keV \cite{BES:02:2})
respectively. At 3.74 GeV open $D\overline{D}$-production starts.
In the continuum 4 more resonances have been identified: a relatively
small resonance $\psi(3.770)$ just above $D\overline{D}$-threshold and three
broader resonances, $\psi(4.040)$, $\psi(4.159)$ and $\psi(4.415)$.
From the data it can be  seen that in the energy range above 4 GeV the
background continuum gives a significant contribution.

Now we want to give a model description for the background. Our motivation
is twofold. As already seen in the last section, it is interesting to
separate the background and resonance contributions to estimate the 
relative size and the importance of the higher resonances. 
Furthermore, in practical applications it is more convenient to deal with
a smooth approximating function in terms of a few resonance parameters than with
a large number of data points. 

The next channel above $D\overline{D}$ starts at 3.88 GeV with $D\overline{D}^*$ 
and  $D^*\overline{D}$-production. At higher energies
open  the $D_s\overline{D_s}$, $D^*\overline{D}^*$,  
$D_s\overline{D_s}^*$, and  $D_s^*\overline{D_s}^*$
channels. Since $D^*$ has three spin directions, the production of
$D\overline{D}^*$ and  $D^*\overline{D}^*$ is enhanced 
compared to $D\overline{D}$-production \cite{rgg:76,le:76,egkly:80,rr:95}.

One could try to parametrise all these contributions by appropriate
form factors. However, neither theory nor experiment provide sufficient information 
to predict these form factors.
As a consequence, this ansatz would depend on many free parameters that had to
be fitted from the used data. So the result would strongly depend
on the data set and could not be generalised.
Therefore we use a different approach based on perturbative QCD
and model the non-resonant background production by
\begin{eqnarray}
  \label{eq:3.a}
	R^{BG}_c(s)&=& \frac{4}{3} \sqrt{1-\frac{M_{thre}^2}{s}}
	\left(1+\frac{M_{thre}^2}{2s}\right)\nn\\
	&&\times\left(1+\frac{\alpha_s(s)}{\pi}+1.5245\left(\frac{\alpha_s(s)}{\pi}\right)^2-
	11.686\left(\frac{\alpha_s(s)}{\pi}\right)^3\right)\,, \nn\\
	M_{thre}&=&M_{thre}^{DD^*}=3.879\ \mbox{GeV}\,.
\end{eqnarray}
For $\alpha_s(s)$ we use the three-loop formula with $\Lambda=313\pm 27\ \mev$ which
corresponds to $\alpha_s(M_Z)=0.1172\pm 0.002$ \cite{pdg:02} and three light flavours.
This background is plotted as a dashed line in fig. \ref{fig:2.a}.
For energies sufficiently above $c\overline{c}$-threshold we expect that
the main process will be hard $c\overline{c}$-production which finally
turns into $D$-mesons with unit probability. 
The higher order corrections in eq. \eqn{eq:3.a} are based on perturbation
theory in the massless limit \cite{c:97} and show the correct high energy
behaviour. For a finite charm quark mass the expansion contains soft gluons
which are interchanged between the quarks at threshold. 
These soft gluon ladders, which can be
resummed by means of NRQCD, lead to the formation of the $\psi$-states and are
responsible for the resonance effects. Thus they must not be included in the background
description. 

It remains the choice of the threshold
$M_{thre}$ which appears in the phase space factor. In a QCD-based picture
it is given by the charm pole mass $M_{thre}=2M_c$. However, in this case we 
describe $D$-meson production, so the phase space should rather be given by the phase
space the $D$-meson than that of the charm mass. As described above,
the production of $D\overline{D}$ is suppressed to  $D\overline{D}^*$. In addition,
for higher energies the other heavier thresholds open. Consequently a 
threshold of $M_{thre}=M^{DD}_{thre}$ would probably overestimate the
background in the intermediate energy range. Therefore we fix the threshold
parameter to $M_{thre}=M_{thre}^{DD^*}=3.879\ \mbox{GeV}$.
One of the limitations of this description is obvious: one cannot expect
it to be a good approximation below and directly above the  
$D\overline{D}^*$-threshold in the range of $3.74\ \gev <\sqrt{s}<4.0\ \gev$. 
Below  $3.88\ \gev$  we obviously miss the small $D\overline{D}$-production
which may also have some impact directly above the $D\overline{D}^*$ threshold.
However, we expect that \eqn{eq:3.a} gives
a good description of the background in the energy region between
$4.0\ \gev <\sqrt{s}<4.6\ \gev$. Above $\sqrt{s}\gsim 4.6\ \gev$ the resonance structure
seems to level off into a continuum. Here one cannot separate 
the background and resonances any longer. Thus, instead of the background, one
should use the full theoretical result to describe the spectral density for
energies above 4.6 GeV. 

Now we give a description of the cross section in terms of the
background and Breit-Wigner resonances:
\begin{equation}
  \label{eq:3.b}
	R_c(s)=R^{BG}_c(s)+\frac{9s}{\alpha^2}\sum_{k=3}^{6}
	\frac{\Gamma_{ee}^{k}\,\Gamma_{tot}^{k}}{(s-E_k^2)^2+E_k^2\,
	\Gamma_{tot}^{k\,2}}\,.
\end{equation}
We prefer to use a constant total width $\Gamma_{tot}^{k}$
instead of a $s$-dependent one since the functional form close to threshold 
is not clear and the uncertainty connected with the background description is
at least of the same order.
With this formula one can extract the resonance parameters from 
the BES data \cite{BES:01} between 3.7 GeV and 4.6 GeV (75 data points). 
The statistical and systematic error have been added quadratically for each
data point. In table \ref{tab:3.a} we have listed our results for 
$E_k$, $\Gamma_{tot}^{k}$ and $\Gamma_{ee}^{k}$ for the resonances
$\psi(3)-\psi(6)$. The fit gives a $\chi^2/dof=1.02$. 
\begin{table}
\begin{center}
\begin{tabular}{|l|l|l|l|}\hline
\multicolumn{1}{|c|}{$k$} &
\multicolumn{1}{|c|}{$E_k/[\gev]$} &
\multicolumn{1}{|c|}{$\Gamma_{tot}^{k}/[\mev]$} &
\multicolumn{1}{|c|}{$\Gamma_{ee}^{k}/[\kev]$} \\ \hline 
3 & $3.7727\pm 0.0016$ & $24.4\pm 4.3$ &  $0.19\pm 0.025$  \\
4 & $4.0504\pm 0.0043$ &  $98.5\pm 12.8$ & $1.03\pm 0.11$  \\
5 & $4.1665\pm 0.0061$ & $55.9\pm 12.3$ & $0.37\pm 0.081$  \\
6 & $4.4294\pm 0.0085$ & $86.0\pm 20.9$ & $0.39\pm 0.074$ \\ \hline
\end{tabular}
\caption{\label{tab:3.a}  
$E_k$, $\Gamma_{tot}^{k}$ and $\Gamma_{ee}^{k}$ for the resonances
$\psi(3)-\psi(6)$ from eq. \eqn{eq:3.b}
and using BES data \cite{BES:01} between 3.7 GeV and 4.6 GeV.}
\end{center}
\end{table}
\begin{figure}
\begin{center}
\vspace{5mm}
\includegraphics[width=\figwidth,height=\figheight,]{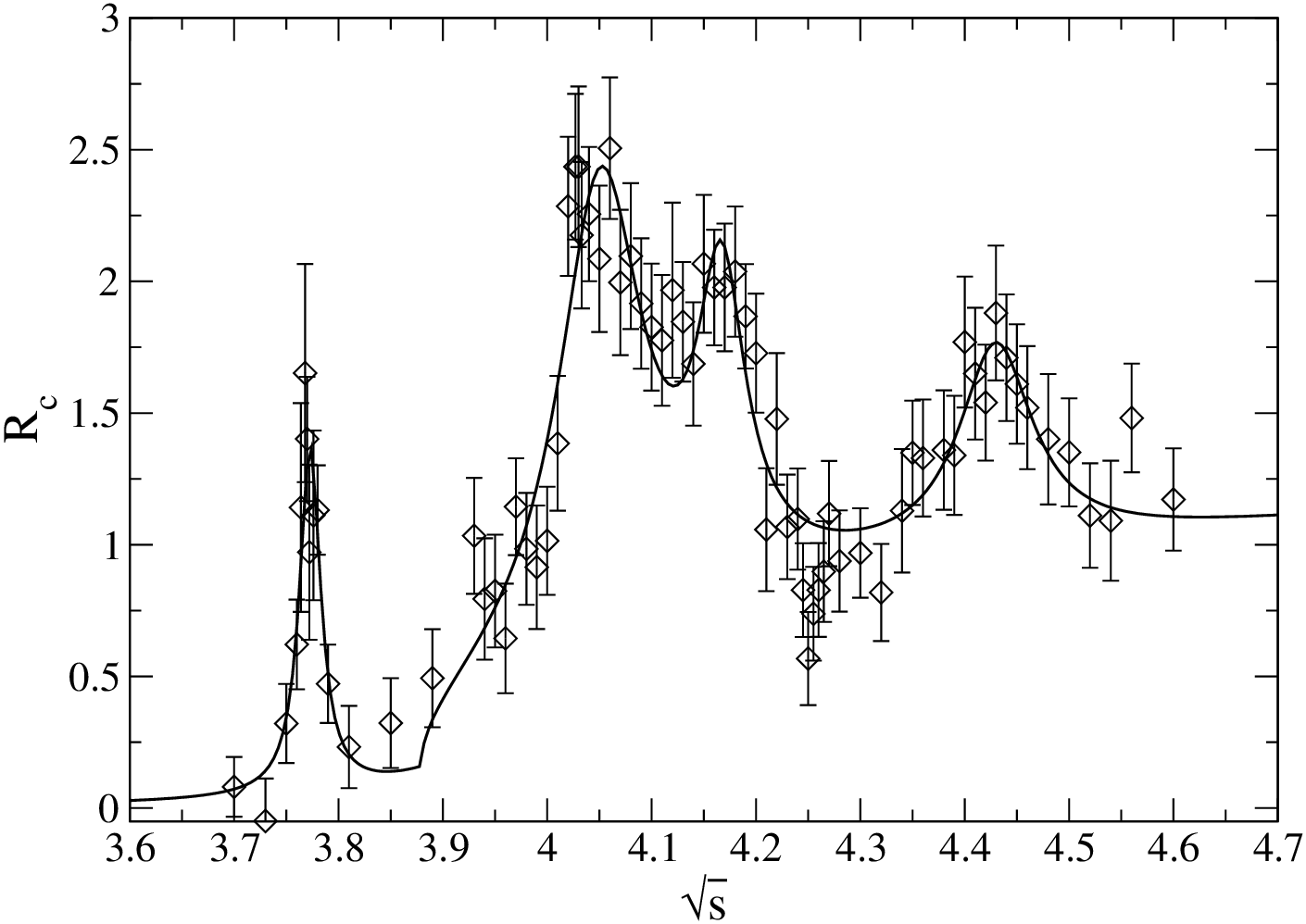}
\caption{\label{fig:3.a} Spectral density from \eqn{eq:3.b} with resonance
parameters from tab. \ref{tab:3.a} and BES data \cite{BES:01}.}
\end{center}
\end{figure}
In fig. \ref{fig:3.a} we have plotted the resulting spectral density together
with the data points. It can be seen that the experimental spectral density is
rather well approximated by the theoretical description. 
We can compare the results to former measurements where the estimate of the
non-resonant $D$-production was fitted to the data 
\cite{betal:78,setal:80,pdg:02}.
We see that the masses change only by several MeV. 
Whereas the change of $\Gamma_{tot}^{k}$ and $\Gamma_{ee}^{k}$ 
for the $\psi(3)$ is mild, these changes are larger for the higher resonances
since these parameters are not only based on a different data set but also on a
different background description.

To investigate closer the dependence of the resonance parameters on the
background we now vary the threshold energy $M_{thre}$. We compare to the results
obtained from a background with $M_{thre}=3.74\ \gev$ which would correspond to
$D\bar{D}$-production and to $M_{thre}=4.0\ \gev$. Compared to the data, the
first one seems to overestimate and the second one to underestimate the
background. 

\begin{table}
\begin{center}
\begin{tabular}{|l|l|l|l|l|l|l|}\hline
\multicolumn{1}{|c|}{} &
\multicolumn{2}{|c|}{$E_k/[\gev]$} &
\multicolumn{2}{|c|}{$\Gamma_{tot}^{k}/[\mev]$} &
\multicolumn{2}{|c|}{$\Gamma_{ee}^{k}/[\kev]$} \\ \hline 
$k$ & \multicolumn{1}{|c|}a & \multicolumn{1}{|c|}b & \multicolumn{1}{|c|}a &
\multicolumn{1}{|c|}b & \multicolumn{1}{|c|}a & \multicolumn{1}{|c|}b\\ \hline 
3 & 3.7715 & 3.7725 & 18.0 & 22.4  & 0.12 & 0.17 \\
4 & 4.0504 & 4.0450 & 75.5 & 138.1 & 0.74 & 1.60 \\
5 & 4.1629 & 4.1701 & 56.4 & 51.5  & 0.36 & 0.32 \\
6 & 4.4296 & 4.4299 & 70.8 & 98.3  & 0.30 & 0.48 \\ \hline
\end{tabular}
\caption{\label{tab:3.b}
Resonance parameters for different backgrounds with $M_{thre}=3.74\ \gev$
(a) and $M_{thre}=4.0\ \gev$ (b). The statistical errors are similar to
table \ref{tab:3.a}.}
\end{center}
\end{table}
We have listed the results in table \ref{tab:3.b}. The statistical
errors are very similar to table \ref{tab:3.a} and have therefore been omitted.
The masses remain very stable. The masses of the $\psi(3)$ and  $\psi(6)$
do not change,  $\psi(5)$ changes by $\Delta E_5=\pm 3.5 \ \mev$ and the largest
change is of the  $\psi(4)$ with  $\Delta E_4=-5 \ \mev$ for $M_{thre}=4.0\ \gev$.
For $M_{thre}=3.74\ \gev$ the widths of the  $\psi(3)$ are reduced since the
background already starts below the resonance. The widths of the  $\psi(5)$
seem to be very stable against variation of the background and the change for
the $\psi(6)$ is of a similar size as the statistical errors. The widths  of
the  $\psi(4)$ are the most sensitive to a variation of the background. They
show a significant change which is clearly larger than the statistical
error. These changes of the resonance parameters are part of the systematic
uncertainty connected with the background description and of the difficulty to
separate these two contributions. However, the experimental cross section is
well approximated by eqs. (\ref{eq:3.a},\ref{eq:3.b}) and the resonance
parameters of table \ref{tab:3.a}.


\section{Theoretical versus phenomenological moments}

QCD sum rules provide a framework which relates a QCD-based description
in terms of QCD parameters to measurable quantities in terms of
hadron properties. In this section we discuss the different 
conceptions related to the theoretical and phenomenological description.

The advantage of using a weight function as in \eqn{eq:2.e} is that
the moments $W_n$ are directly connected with $\Pi(s)$
at the Euclidean point $s=-\Delta$:
\begin{equation}
  \label{eq:4.a}
	W_n(\Delta)=W_n(\Delta,s_{min},\infty)=
	\frac{12\pi^2}{n!}\frac{d^n}{ds^n}\Pi(s)\Big|_{s=-\Delta}\,,
\end{equation}
where $s_{min}$ indicates the lowest pole.
In the Euclidean region the theoretical expansion is known to be valid
by means of the OPE. In addition to the perturbative result, condensates
of higher and higher power will appear. 
In a QCD-based picture the definition of the pole mass provides the natural 
description for the onset of the continuous spectral density and
therefore the threshold is given by $\sqrt{s}=2M_c$. 
The theoretical expansion depends on the values of $\Delta$
and $n$: large values of $\Delta$ and small $n$ move the moments to a safe
perturbative region and the expansion in $\alpha_s$ converges well. In
principle, $W_n$ could thus be calculated to high accuracy. However, this
region is of little phenomenological interest. Usually in sum rules analyses
one is interested in extracting information on the ground state or the quark
masses. In order to be sensitive to these parameters the analysis must be
performed relatively close to threshold. In this case the perturbative
expansion does not converge well any more since large terms appear reflecting
the Coulombic structure of the charmonium system. These potentially
large terms can be resummed with the method of NRQCD which sets up a systematic
framework to treat these non-relativistic corrections. The result is expressed
in terms of a Green's function and can be directly evaluated at
$s=-\Delta$. Its imaginary part shows poles below and a continuum spectral
density above threshold.

This QCD-based theoretical description has to be confronted to the
measured cross section to which it is related by the optical theorem.
As described in the last section, its behaviour is very complicated: 
It contains two sharp resonances and several higher resonances which are
shifted into the continuum.
It is obvious that the theoretical and phenomenological spectral density 
do not equal each other. In the Euclidean region the theoretical expansion is
truncated -- as a series in $\alpha_s$ and in higher condensates. If one were
able to calculate $\Pi(q^2)$ in the Euclidean domain exactly, one could
analytically continue the result to the Minkowski domain and take the imaginary
part. The theoretical spectral density would equal the hadronic cross
section. However, in practice only the truncated expansion is analytically
continued to values of positive $q^2$. Small neglected terms in the
Euclidean domain can become large in the Minkowski region and change the
spectral density significantly. The origin and
behaviour of such contributions have been discussed in \cite{s:00}.

In the OPE condensates appear, the leading contribution is given by
the gluon condensate:
\begin{equation}
  \label{eq:4.b}
	\Pi_{FF}(s)=\frac{4}{9}\frac{\langle \frac{\alpha_s}{\pi}FF\rangle}{16 M_c^4}
	\left(C_{FF}^{(0)}(s)+\frac{\alpha_s}{\pi}C_{FF}^{(1)}(s)\right)\,,
\end{equation}
the analytic form of the functions $C_{FF}^{(0)}$ and $C_{FF}^{(1)}$
can be found in \cite{bbifts:94}. However, 
its contribution to the moments is small for values of $n$ and $\Delta$
used in this analysis. Its contribution grows if one comes very close
to threshold, for $\Delta\gsim -4M_c^2$ or for very large $n$ since here the
moments test the nonperturbative region.

Now we want to compare the size of the individual theoretical and phenomenological
contributions. In fact, this comparison is done in QCD sum rules to extract the
charm and bottom quark masses since the moments show a strong dependence on the
value of the mass.
We fix the ${\rm \MSb}$-mass to $m_c(m_c)=1.19\ \gev$ \cite{e:02}.
For the comparison we use a range of values for $n$ and $\Delta$ somewhat
larger than in typical sum rule applications.
In table \ref{tab:4.a} we show the results for the theoretical and experimental
moments.
\begin{table}
\begin{center}
\begin{tabular}{|c|c|c|c|}\hline
\multicolumn{4}{|c|}{$\Delta=0$} \\ \hline 
$n$ & $3$ & $7$ & $12$ \\ \hline
Theory total & 1.04 & 0.85 & 0.67 \\ 
Theory poles & 0.71 & 0.76 & 0.65 \\ 
Theory continuum & 0.33 & 0.086 & 0.021 \\ \hline
Exp. total     & 1 & 1 & 1 \\ 
Exp. poles 1+2 & 0.87 & 0.989 & 0.9992 \\ 
Exp. BES data  & 0.13 & 0.011 & 0.0008 \\ \hline\hline
\multicolumn{4}{|c|}{$\Delta=(4\ \gev)^2$} \\ \hline 
$n$ & $3$ & $7$ & $12$ \\ \hline
Theory total & 0.98 & 1.10 & 0.99 \\ 
Theory poles & 0.34 & 0.72 & 0.80 \\ 
Theory continuum & 0.64 & 0.38 & 0.19 \\ \hline
Exp. total     & 1 & 1 & 1 \\ 
Exp. poles 1+2 & 0.56 & 0.86 & 0.96 \\ 
Exp. BES data  & 0.44 & 0.14 & 0.04 \\ \hline
\end{tabular}
\caption{\label{tab:4.a}  
Charmonium: Theoretical and phenomenological moments normalised to the total phenomenological
moments.}
\end{center}
\end{table}
The moments have been normalised to the total phenomenological
moments. The poles represent the dominant contribution and they are even more
pronounced on the phenomenological side. It can be clearly seen that small
$\Delta$ and large $n$ shift the analysis closer to the poles. For $n=12$ the
continuum region is essentially cut off. For this value of the charm mass the moments
show a good stability, which is no surprise since this stability criterion 
was used to extract the
value of the charm mass \cite{e:02}. The total theoretical moments differ from the
phenomenological ones by about 10\%. The convergence is better for large values
of $\Delta$. Only for $\Delta=0$ and $n=12$ they differ significantly, but here
the theoretical moments are evaluated close to threshold and one cannot expect
a reliable description of these moments.

It is interesting to
investigate the dependence
of the theoretical moments on the mass. With a ${\rm \MSb}$-mass of
$m_c(m_c)=1.3\ \gev$ the total theoretical moments normalised to the
phenomenological ones for $\Delta=0$ vary from 0.82 (for $n=3$) to 0.21
($n=12$) and from 0.96 ($n=3$) to 0.68 ($n=12$) at $\Delta=(4\ \gev)^2$.
For $m_c(m_c)=1.1\ \gev$ the relative theoretical moments for $\Delta=0$ increase
from 1.33 ($n=3$) to 2.27 
($n=12$) and from 0.97 ($n=3$) to 1.4 ($n=12$) at $\Delta=(4\ \gev)^2$.
It can be seen that the sensitivity of the moments on the mass decreases for
large values of $\Delta$.
The change is due to the pole contribution since the continuum spectral density
is practically independent of the mass. These results confirm the 
significant effect of the mass on the moments and on the stability of the sum rules.


\section{Upsilon system}

In the upsilon system the experimental situation is unsatisfactory. Apart from
the resonance parameters of the 6 $\Upsilon$-states \cite{pdg:02} almost no
direct information on the cross section above the $\Upsilon(4)$ is
available. Therefore it is not possible to perform a check of QHD as has been
done in section 2. However, we can use the results of the last sections to draw
some conclusions for this energy region.

First we compare the size of resonance and background contribution.
Analog to eq. \eqn{eq:3.a} we make the following ansatz:
\begin{eqnarray}
  \label{eq:5.a}
	R^{BG}_b(s)&=& \frac{1}{3} \sqrt{1-\frac{M_{thre}^2}{s}}
	\left(1+\frac{M_{thre}^2}{2s}\right)\nn\\
	&&\times\left(1+\frac{\alpha_s(s)}{\pi}+1.4092\left(\frac{\alpha_s(s)}{\pi}\right)^2-
	12.805\left(\frac{\alpha_s(s)}{\pi}\right)^3\right)\,, \nn\\
	M_{thre}&=&M_{thre}^{BB^*}=10.604\ \mbox{GeV}\,.
\end{eqnarray}
The strong coupling constant at three loop with 4 light flavours
is determined from  $\Lambda=278\pm 30\ \mev$. Similar to the charm case,
we fix the start of the threshold at the energy for $BB^*$-production.

We can compare the background to the resonance contribution. 
\begin{figure}
\begin{center}
\vspace{5mm}
\includegraphics[width=\figwidth,height=\figheight,angle=-0]{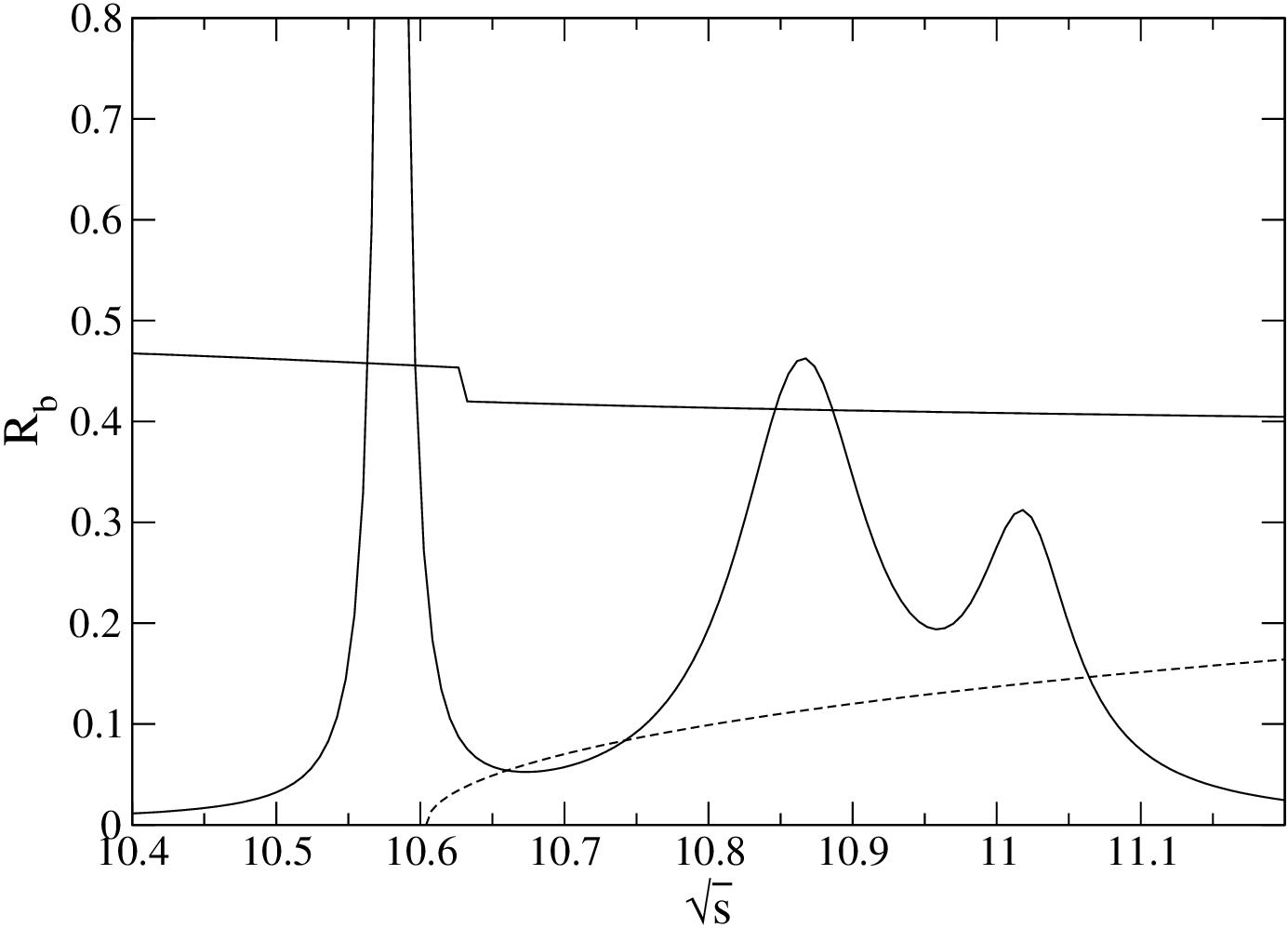}
\caption{\label{fig:5.a}
$\Upsilon(4)-\Upsilon(6)$ with theoretical spectral density (solid)
and background (dashed).}
\end{center}
\end{figure}
\begin{table}
\begin{center}
\begin{tabular}{|c|c|c|c|}\hline
\multicolumn{4}{|c|}{$\Delta=0$} \\ \hline 
$n$ & $3$ & $7$ & $12$ \\ \hline
Theory total & 1.05 & 0.94 & 0.76 \\ 
Theory poles & 0.41 & 0.60 & 0.61 \\ 
Theory continuum & 0.64 & 0.34 & 0.15 \\ \hline
Exp. total     & 1 & 1 & 1 \\ 
Exp. poles 1+2 & 0.70 & 0.94 & 0.991 \\ 
Exp. BES data  & 0.30 & 0.06 & 0.009 \\ \hline\hline
\multicolumn{4}{|c|}{$\Delta=(10\ \gev)^2$} \\ \hline 
$n$ & $3$ & $7$ & $12$ \\ \hline
Theory total & 1.04 & 1.06 & 0.99 \\ 
Theory poles & 0.22 & 0.45 & 0.58 \\ 
Theory continuum & 0.82 & 0.61 & 0.41 \\ \hline
Exp. total     & 1 & 1 & 1 \\ 
Exp. poles 1+2 & 0.45 & 0.75 & 0.91 \\ 
Exp. BES data  & 0.55 & 0.25 & 0.09 \\ \hline
\end{tabular}
\caption{\label{tab:5.a}  
Upsilon: Theoretical and phenomenological moments normalised to the total phenomenological
moments.}
\end{center}
\end{table}
In fig. \ref{fig:5.a} we have plotted the background versus the spectral
density using the resonance parameters $\Upsilon(4)-\Upsilon(6)$ from \cite{pdg:02}.
The data indicate \cite{b:85,l:85}
that in the region between $11.0 \ \gev<\sqrt{s_0}<11.2\
\gev$ the cross section continues at values around the peak of the
$\Upsilon(6)$. It can be seen that
the resonance contribution exceeds the background contribution in the range
until $\sim 11.05$ GeV.
In the resonance region the influence of the background is not as strong
as in the charmonium system. Above 11.05 GeV, since no further resonances
are measured, one should use the full theoretical spectral density to describe 
the cross section.

An important parameter in sum rule calculations of the bottom mass is the
threshold parameter $s_0$. The phenomenological moments are determined from
the first six $\Upsilon$-resonances and the theoretical spectral density above
$s_0$. With no background production this parameter would be
estimated to lie about 250 MeV above the $\Upsilon(6)$. However, the
non-resonant background, where it is not already included in the resonance
parameters, will effectively lower this threshold.
In chapter 2 two ways to estimate this parameter have been discussed, 
by a model description for the background 
and by assuming QHD already for states $n> 3$. In the first one 
we use the background of eq. \eqn{eq:5.a}. 
In the region between $10.75-11.05\ \gev$ the total cross section
is well approximated by the $5^{th}$ and $6^{th}$ resonance. 
Calculating the background for the remaining interval the effective threshold
lowers to $\sqrt{s_0}=11.15-11.20\ \gev$.
In the second approach we assume QHD for states with $n>3$.
The background from this ansatz would reduce the threshold to  
$\sqrt{s_0}=11.0-11.05\ \gev$.
Therefore, taking the average between both estimates, we suggest that a
reasonable estimate for $s_0$ would be given by $\sqrt{s_0}=11.1\pm 0.2\ \gev$.

As in section 4 we can compare the contributions from the theoretical and
phenomenological moments. Again we evaluate the moments
in a region where threshold effects are important.
In table \ref{tab:5.a} we have collected the moments from the different
sources. As input parameter for the mass we have used $m_b(m_b)=4.24\ \gev$ \cite{e:02}.
For the employed values of $n$ and $\Delta$ the dominance of the poles is less
pronounced than in the charmonium. The theoretical moments differ by about 5\%
from the phenomenological ones. Thus they show a better convergence than in the
charmonium system. At $\Delta=0$ and $n=12$ the theoretical calculation of the
moments should not be trusted any more.

Now we look at the influence of the bottom mass on the
moments. With a ${\rm \MSb}$-mass of $m_b(m_b)=4.34\ \gev$ the total
theoretical moments for $\Delta=0$ vary 
from 0.93 (for $n=3$) to 0.47 
($n=12$) and from 0.99 ($n=3$) to 0.80 ($n=12$) at $\Delta=(10\ \gev)^2$.
For $m_b(m_b)=4.14\ \gev$ the moments for $\Delta=0$ increase
from 1.19 ($n=3$) to 1.23 
($n=12$) and from 1.1 ($n=3$) to 1.24 ($n=12$) at $\Delta=(10\ \gev)^2$
thus showing again the strong influence of the quark mass.


\section{Conclusions}

In this work we have explored the cross section of $e^+ e^-$--collisions
in the charmonium and upsilon energy region. The experimental situation for the
charmonium has significantly improved with the new results from BES
\cite{BES:01}. This allows a thorough comparison of the theoretical and
experimental description for this energy range. 
These investigations have been applied to heavy-heavy systems. Apart from
the general discussions it is not clear if the concrete results could be 
generalised to e.g. heavy-light or light-light
states since the underlying physical systems are different and their properties
are rather determined from the dynamics of the light quarks.

The main part of the paper has investigated the charmonium energy range. We have
given a prescription for the threshold parameter $s_0$ which is needed to describe
the experimental spectral density for large energies. The error on this
quantity was estimated to $\Delta \sqrt{s_0}=200\ \mev$ which includes a
possible variation of $s_0$ with $n$ and $\Delta$. The phenomenological cross
section can be described as Breit-Wigner resonances and a non-resonant
background production of $D^{(*)}$-mesons. We have presented a model for this
background production based on perturbative QCD. The masses, the hadronic
widths and the partial $e^+e^-$--widths of the states $\psi(3)-\psi(6)$ have
been extracted from the BES data. So we obtain a direct theoretical
description of the experimental cross section in terms of few resonance
parameters with a $\chi^2/dof=1.02$.
QCD sum rules use the identity of the theoretical and phenomenological moments
related by the optical theorem and can be used to extract e.g. the
quark masses or the ground state properties. In order to be
sensitive to these parameters the moments must be evaluated in a region not too
far from threshold. Thus the contributions from NRQCD form an essential part on
the theoretical side reflecting the fact that the underlying system is a
Coulombic one. The different contributions from the poles and the continuum
part have been compared for different values of $n$ and $\Delta$.

Section 5 has been devoted to the upsilon system. Here the experimental
situation is dissatisfactory. However, the results of the previous sections
could be used to investigate several properties of this system: a model
description of non-resonant $B^{(*)}$-production has been presented and
the threshold parameter $s_0$ estimated.
We have compared the different contributions to the theoretical and
phenomenological moments and investigated the effect of the bottom mass. 
Unfortunately the cross section is not well measured in the
region above the $\Upsilon(4)$. A more detailed knowledge would allow a better
test of QHD in this energy region. 

The two basic pictures of QCD are related by
QHD: the hadronic world and a description based on perturbative QCD in terms
of quarks and gluons. Thus a better understanding of QHD could provide further insight
into the structure and behaviour of nonperturbative contributions.


\bigskip \noindent
{\bf Acknowledgements}

\noindent
I would like to thank Antonio Pich for interesting 
discussions and reading the manuscript. 
This work has been supported in part by TMR, EC
contract No. RTN2-2001-00199, by MCYT (Spain) under grant
FPA2001-3031, and by ERDF funds from the European Commission.
I thank the Deutsche Forschungsgemeinschaft for financial support.


\end{document}